# A Search for Gamma-Ray Emission from Variable or Transient Galactic Radio Sources

C.R. Shrader
*Astrophysics Science Division, NASA GSFC Code 661, Greenbelt, MD, 2077 USA*
also with Universities Space Research Association

D.J. Macomb
*Boise State University, Dept. Of Physics, 1910 Univ. Drive Boise, Boise ID 83725 USA*

(Abstract) We describe a recent effort using the *Fermi* Large Area Telescope to search for gamma-ray emission from a source sample derived from published surveys of variable or transient galactic radio sources. We are also making follow up searches for hard-X-ray emission using archival data from *Swift* and *INTEGRAL*. We describe our sample selection, data extraction and analysis methods and present results obtained to date. Radio transients could, for example, with high-mass X-ray binaries similar to LS I+61 303, but at lower intensities. We've searched about 45 source fields with *Fermi*, using multi epoch analyses – days to weeks – as well as approximately 12000 data sequences (~$10^2$ to $10^3$-s duration) with *Swift*. We have also constructed INTEGRAL light curves for the 10 best Fermi candidates. This led to 10 (*>200-MeV*) *Fermi* candidates, or and several marginal candidate hard-X-ray transients. We speculate on possible non-gamma-ray source associations and on potential source populations that could be revealed through our analysis.

## 1. INTRODUCTION

Radio emission and in particular variable emission was established early on as a ubiquitous property of point-source of gamma ray emission prior to the launch of the Fermi Gamma Ray Space Telescope (*Fermi*), e.g. [1]. Particle acceleration is widely believed to underlie this synergy. The recent compilation of the second Fermi all-sky source catalog, with its nearly 1900 sources further supports this idea, and a number of gamma-ray – radio source populations have been established. While the most prominent examples involve the radio-loud AGN, spin-powered pulsars and super-nova remnants (SNRs) much less is known about other galactic populations. Regular transient searches are routinely carried out by the Fermi team, but these are relatively bright ($> 10^{-6}$ ph/cm^2/s) at arbitrary locations with typical time scales of hours to days. Our study, on the other hand, examines the specific locations associated with the radio sources to the limiting sensitivity using one-to-three week long accumulations. Yet another approach uses time-domain searches for periodic signals [18]. That has led to the identification of several candidate gamma-ray bright binaries, however, it would be insensitive to non-periodic phenomena and relatively insensitive to periodic phenomena superposed on long-term transient flux histories.

Several galactic populations have been well established gamma ray emitters. Radio loud AGN comprise the predominant population among Fermi sources on the sky and the ISM is transparent at ~GeV energies. Pulsars are a prolific galactic population that are steady radio and gamma-ray emitters on long (greater than tens of seconds) timescales, although they of course exhibit rapid, periodic variations associated with the rapid rotation of the magnetospheres relative to our line of sight. There are also radio-quiet pulsars, but these may just represent cases where radio emission is not seen due to viewing geometry. For a recent review see [17]. The pulsars also have a fairly distinct spectral characteristic in the gamma-ray domain, with the precipitous spectral breaks at a few GeV.

In addition to pulsars and radio-loud AGN, some new classes of point-like gamma-ray emitters have emerged in the *Fermi* era, and again, time-variable radio emission is common characteristic. These newly discovered classes of gamma-ray emitters include counterparts to high-mass X-ray binaries such as Cyg X-3, LSI +61 303 and LS5039 [6,7,8,9] and are commonly referred to as "gamma-ray binaries." This relatively small sub-sample of the galactic X-ray binary population, which as noted are additionally characterized by variable or transient radio emission, have been established as gamma-ray emitters on the basis of temporal domain study, e.g. [18]. Excess Fourier power at discrete frequencies that coincide with known orbital periods are the definitive signature. In addition, aperiodic and episodic variations are also observed, and some patterns in their multi-wavelength behavior have begun to emerge.

The gamma-ray binaries are also often transient X-ray emitters and since they tend to reside in highly obscured regions of the galactic plane they are often revealed through survey observations in the hard-X-ray band. The galactic plane coverage of the INTEGRAL hard-X-ray telescope (IBIS) in particular has significantly impacted our accounting of the high-mass X-ray binary population. The mechanism through which some of these objects emit gamma-rays is unclear. Particle acceleration leading to a jet or a jet-wind or wind-wind shock region (*e.g.* analogous to the pulsar wind nebulae, but contained within the circum-stellar environment) are viable possibilities. Our survey has the potential to reveal additional objects among this class, establishing them as γ-ray emitters and thus help to address this issue.

There are a number of other possible source classes that could be associated with variable radio emission or radio transients. For some of these, one would not expect to find gamma-ray emission whereas for others it would be less surprising. Possible radio transient source classes might include magnetars, RRATs (Rotating Radio Transients), classical novae and coronally active stars (particularly those in in binaries), or flare stars. The latter would not be expected to be visible to Fermi, even though flares can exceed those seen in the sun by orders of magnitude. Of course, and perhaps most interestingly, totally unexpected phenomena could also be revealed.

In this paper we present results from our search for transient or variable hard X-ray and gamma-ray emission from a candidate sample of Galactic plane radio transient and/or high-amplitude variables. We describe or sample selection, data analysis approach and identify several promising candidates. We discuss possible non-gamma-ray counterparts to our best candidate detections and consider the possible implications on source populations. We describe or source sample selection criteria in section 2. In section 3 we describe our data analysis approach and we then present the results of that analysis in section 4. Discussion and conclusions are offered in section 5.

## 2. SAMPLE SELECTION

While radio variability is often associated with of high-energy processes and can lead to the discovery rare or enigmatic objects the required observational campaigns are in practice relatively rare. This is due primarily to their resource-intensive nature combined with high proposal pressure on the suitable telescope facilities. The length of time needed to carry out reasonable campaigns is generally longer than observatory allocation cycles. There are some efforts of note in the Galactic domain however. Extensive multi-epoch galactic-plane survey work was first carried out in the 1980s by Gregory and Taylor (1986; 1981). Those surveys covered about 20% of the galactic plane at *6 cm* with typically several arcminute spatial resolution. More than 1200 source were cataloged over the course of a 5-year campaign. Those authors identified about 36 optimal variable-flux candidates on timescales of interest to our study. Some unknown fraction of these may still be active and may be associated with galactic rather than background extragalactic counterparts.

More recently several new radio surveys of portions of the galactic plane based on multi-epoch observations have been published, notably, the recent work of [2]. That survey covers 3 epochs over a *~15-year* time span, covering ~23 *sq*. deg, about *±1°*in latitude, to *~100 mJy* at *6 cm*. Those authors also present IR, optical and X-ray counterpart searches for their sample objects. They present and arguments against contamination of their sample by background AGN or scintillation effects involving the galactic pulsar population. This tends then to point towards association with XRBs, radio-loud magnetars, or some unknown class of radio transients. This survey was a primary resource for our work. We focused on cases with both a high (>10) ratio of maximum-to-minimum flux excursions and a minimum peak flux of 10 mJy or greater. This resulted in 26 objects.

We also have searched the variable source candidates identified in the earlier Galactic plane survey of [11,12]. That survey covered ~20% of the galaxy at *6 cm*, identifying ~35 optimal candidates. Among those 35 sources, we've thus far included about half optimized on again observed variability amplitude (>10) OR a minimum flux >100 mJy with at least a 1.5 flux range. In addition we investigated the survey work of [10], although their sample objects tend to lie off the galactic plane by 5-6 degrees. We ultimately did not identify any compelling candidates from this sample so subsequent discussion will focus on the other two.

There are some potentially significant difficulties associated with these types of surveys [2]. Radio signals can be modified by the effects of Galactic scintillations which would result in spurious transient or variable candidates. Thus some of the sample objects could be in actuality, for example, constant intensity pulsars or SNRs. The spectral energy distribution, both in radio and gamma ray can help screen against the former possibility. One must also consider the likely contribution; maybe even the predominance, to ones sample of background extragalactic sources. These would be the radio loud quasars to which the Galactic plane is transparent at GHz frequencies as well as in GeV gamma rays. Accepting the arguments of [2] that candidate list comprises a sample with a relatively low probability of extra-galactic associations (based on variability amplitudes, longitude distribution and spectra), and which excludes pulsars (which could appear variable due to ISM scintillations, but have distinct spectra). Coronally active stars are also possible associations, but there is a dearth of bright visual counterparts at least for the Becker et al. (2010) [2] sample. Other possible counterparts are high-mass X-ray binaries analogous to the previously mentioned Cyg X-3 or LSI +61 303, or perhaps like or SS 433. Radio loud magnetars [19], and putative objects resembling an unidentified high-amplitude transient in the galactic-center region [14] are all potential associations as well as being potential γ-ray emitters.

## 3. DATA ANALYSIS

### 3.1. Fermi

Once we defined our list of source candidates we performed multi-epoch analyses using data from the first 4.5 years of data obtained with the Large Area Telescope (LAT) experiment onboard *Fermi*. The standard point-source detection and flux determination method is based on a maximum-likelihood (ML) analysis [15]. Following the approach outlined therein we used the *Fermi* science tools to derive maximum-likelihood flux determinations, or flux upper limits for each source at multiple epochs spanning the first 4.5 years of the mission. The typical time intervals searched were 6, 12 and 18-day at uniformly placed epochs spanning that 4.5-year baseline.

Sky models for regions of interest were derived from the 2FGL catalog [5] combined with the Galactic and isotropic background models distributed with the *Fermi* science tools up to date as of late 2012. Regions of interest of about 15 degrees were included in our analysis, but with source-model parameters typically frozen to the 2FGL catalog values for annular regions at radii 10 degrees or greater. The analysis chain was scripted to facilitate processing of the large data volume and parameter space. Our most promising candidates were subsequently followed up by constructing epoch specific significance maps of the region. We also probed different timescales and considered possible source position refinements based on the gamma-ray analysis.

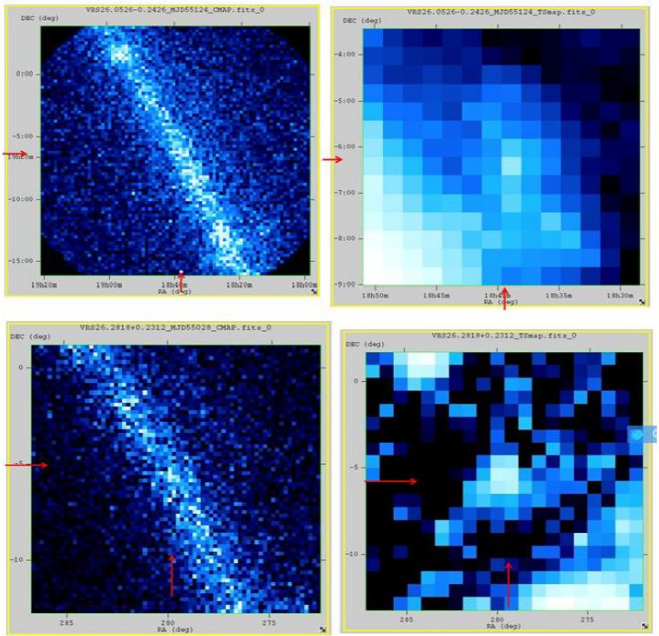

Figure 1. Examples of count maps (lhs) and TS maps (rhs) for two flare candidates from our sample derived from the *Fermi* analysis as described in the text.

We also attempted an alternative analysis using the aperture-photometry method [18], which is less accurate as backgrounds are not modelled or subtracted, but on the other hand is substantially less computationally demanding. We found however, that while that method may suffice for temporal-domain searches for periodic phenomena at moderate intensity, it was of limited value in our case, as we required searches to deep limiting flux. Thus, the ML approach was our method of choice for our definitive analysis.

**eConf C121028**

### 3.2 Swift

We continue to monitor the Swift/BAT archives for data covering our survey fields–. Estimates of flux and detection significance for individual BAT scans for the sources of [2] from 2005 through 2012 is complete. Accumulating a similar dataset for sources from [10] is mostly complete, and data for sources from [11,12] is in progress.. Our basic approach to the analysis utilized the ftool *"batsurvey"* in 8 energy bands for individual spacecraft orbits. Individual orbit data are then binned into non-overlapping seven day segments. This results in typical exposure time/source of about 5 *Msec* on our survey field.

The base reference catalog used for data reduction was the BAT monitored survey sources found at [16]. Other sources added to the analysis were our variable radio source sample, and the Swift team defined field points which are used for calibration and self-consistency checking. From the full catalog, we proceeded by removing sources co-located within 2 arcminutes. This results in an original 31 source candidates from [2] and 34 from [10] that are sufficiently free from source confusion. The 35 significantly variable sources from have recently been added.

### 3.3 INTEGRAL

For each candidate Fermi detection we also considered the possibility that there may be of hard-X-ray counterparts among the INTEGRAL public data archives. We analysed the IBIS/ISGRI public data from all relevant spacecraft pointings, or "Science Windows" obtained between December 2002 and May 2011 for which the field of interest was in the fully-coded instrument field of view; approximately a 10-degree radius about the telescope bore sight. The INTEGRAL Data reduction was performed using the standard OSA-10 analysis software package available from the INTEGRAL Science Data Center (ISDC).

We also supplemented this analysis by examining quick-look light curves generated using the HEAVENS facility of the INTEGRAL Science Data Center. When using this approach, since it involves far less computation, we included source candidates regardless of whether they had Fermi detection or not. Proceedings.

### 4. RESULTS

Our Fermi analysis yielded a number of possible single-epoch *Fermi* detections, based on the test-statistic threshold of *TS=25*, which is in accordance with recommended procedures of the *Fermi* Science Support Center. We included cases where this test statistic threshold was achieved for at least one epoch for at least one of our time-binning schemes. In this initial assessment we do not consider the number of statistical "trials". This led to a total of 9 ~single-epoch candidate >200-MeV gamma-ray candidates. A compilation of our most compelling transient/variable source candidates are presented in Table 1.

Table 1. Candidate gamma-ray flare detections among our radio-selected source sample.

| Radio Cat ID | RA (deg) | Dec (deg) | l | b | $S_{max}/S_{min}$ | $S_{max}$ | Max TS | Epoch (maxTS) | 2FGL Coincidence? | offset (deg) | Comments |
|---|---|---|---|---|---|---|---|---|---|---|---|
| 21.6552-0.3611 | 277.99 | -10.19 | 21.66 | -0.36 | 66.6 | 67.8 | 41.0 | 55076 | | | X-ray source? ( unidentified XMM. Fx~10^-14) |
| 22.9116-0.2878 | 278.51 | -9.04 | 22.91 | -0.29 | 15.4 | 46.2 | 42.1 | 54956 | 2FGL J1834.3-0848 | 0.24 | 90 cm |
| 22.9743-0.3920 | 278.63 | -9.03 | 22.97 | -0.39 | 13.3 | 14.7 | 25.3 | 54956 | 2FGL J1834.3-0848 | 0.23 | |
| 23.6644-0.0372 | 278.64 | -8.26 | 23.66 | -0.04 | 55.5 | 26 | 26.5 | 55328 | | | |
| 23.5585-0.3241 | 278.85 | -8.48 | 23.56 | -0.32 | 12.4 | 15.6 | 30.6 | 55664 | | | |
| 26.2818+0.2312 | 279.61 | -5.81 | 26.28 | 0.23 | 14.5 | 22.4 | 36.4 | 55028 | 2FGL J1839.0-0539 | 0.22 | ??==1FGL J1839.1-0543c ?? |
| 26.0526-0.2426 | 279.93 | -6.23 | 26.05 | -0.24 | 12.2 | 27.5 | 20.0 | 55124 | 2FGL J1839.3-0558c | 0.27 | |
| 29.7161-0.3178 | 281.68 | -3.01 | 29.72 | -0.32 | 41.2 | 30.4 | 34.1 | 54944 | | | G29.7-0.3 SNR? (6' offset) |
| 30.4376-0.2062 | 281.91 | -2.31 | 30.44 | -0.21 | 38.1 | 17.5 | 42.7 | 55676 | | | ASCA source?, AX J184741-0219 (<5σ) 1' offset |
| 37.2324-0.0356 | 284.86 | 3.81 | 37.23 | -0.04 | 5.8 | 14.7 | 26.9 | 55544 | | | |
| GT0005+626 | 1.97 | 62.94 | 118.00 | 0.49 | 1.6 | 95.00 | 18.0 | 55616 | | | |
| GT0556+238 | 87.18 | 26.59 | 182.36 | -0.63 | 8.2 | 867.0 | 23.3 | 55004 | | | |
| GT0558+234 | 90.44 | 23.41 | 186.60 | 0.32 | 1.5 | 290.0 | 21.5 | 55004 | | | |
| GT0629+103 | 98.05 | 10.36 | 201.53 | 0.49 | 1.4 | 892.00 | 19.1 | 55460 | | | |
| GT2100+468 | 315.57 | 47.04 | 87.94 | 0.34 | 3.1 | 444.00 | 23.1 | 55712 | 2FGL J2103.4+4706 | 0.21 | |

For our best gamma-ray counterpart candidates we considered possible associations with cataloged sources at other wavelengths or with other gamma-ray source catalogs. We obtained various IR-γ-ray point-source catalogs from the HEASARC, as tabulated in Table 2. The criterion applied was a simple spatial coincidence with a search radius equal to the approximate 1-GeV Fermi point-spread function.

Several cases of interest are noted in table 1. In a few cases, for example 21.6552-0.3611 we find an association with a faint ($F_x$ ~$10^{-14}$ ergs/cm$^2$/s) X-ray source, XGPS-I J183157-101123, from the XMM Galactic Plane Survey Catalog.. Depending on galactic absorption on these particular lines of sight, which is likely to be substantial (a few time $10^{22}$cm$^{-2}$ based on the HEASARC NH calculator which is very approximate), these X-ray sources could be consistent with the intensity expected from one of the known γ-ray binaries located at ~kpc distance.

One source, 29.7161-0.3178, is likely associated with an SNR, thus unlikely to be variable However, gamma-ray emission associated with a young and evolving pulsar – e.g. in a magnetar phase –are possible scenarios for transient radio and gamma emitters.

## 5. DISCUSSION AND CONCLUSIONS

We have performed multi-epoch point-source analyses of *Fermi, INTEGRAL* and *Swift* fields centered on a sample of Galactic radio sources culled from the literature. Each of these radio sources had been identified as a possible variable or transient. So far we've analyzed about 45 fields with *Fermi* and *Swift* and a subset of those with *INTEGRAL* where we focused on the most viable Fermi candidates. This is an ongoing work in progress as new data is continuously being collected, and we are currently expanding our sample with other radio monitoring surveys. About 20% of the sources are detected at least one epoch at *TS>~25* (corresponding to *5σ*) significance in our Fermi analysis. However, considering the large number of "trials" associated with a multi-epoch analysis such as the one we have carried out, it is probably more reasonable to set a threshold criteria of ~6-sigma. About 3 of these are likely associated with *2FGL* sources, as they are within the nominal LAT 1-GeV PSF 68%-containment radius.

The *Swift* & *INTEGRAL* analysis has yielded mostly null results. There are several marginal incidences of intensity increases coincident with our *Fermi* flare candidates. We note that by analogy, the hard X-ray signatures of *LS I+61 303* and *LS 5039* are weak and rather non-descript. If those objects were at greater distances and/or viewed through higher galactic column densities only a small number of flare peaks would sporadically appear. A fainter population of similar HMXBs could elude *Swift* or *INTEGRAL* detection without contemporaneous observation at other wavelengths; notably in radio or gamma-ray where the ISM is most transparent. Cygnus X-3 on the other hand exhibits a *Swift-Fermi* behavior for which correlation patterns have been identified. Thus, if our candidates remain viable after further analysis, speculatively they may be fainter examples of LS I+61303. Our analysis is on-going as data continue to be accumulated. Contemporaneous radio-γ-ray variations would ultimately provide the most robust counterpart identification.

This is a work in progress, as new data are being continuously accumulated. There are also new low-frequency radio facilities expected to come online in the near future which should provide a much improved inventory.

Table 2. Multi-wavelength source catalogs cross-correlated with our candidate gamma-ray transients.

| Catalog and Survey Cross Referenced |
|---|
| Second Fermi LATCatalog |
| Third Egret Catalog |
| 4th INTEGRAL Catalog |
| ASCA Galactic Plane Survey |
| Chandra "ChampPlane" |
| HESS TeV Catalog |
| HMXBs Catalog |
| MAGPIS Galactic 20cm Survey |
| Parkes PSR VCat |
| VLA Galactic Plane 6cm Survey |
| WR CatCatalog |
| XMM Calactic Plane Survey |

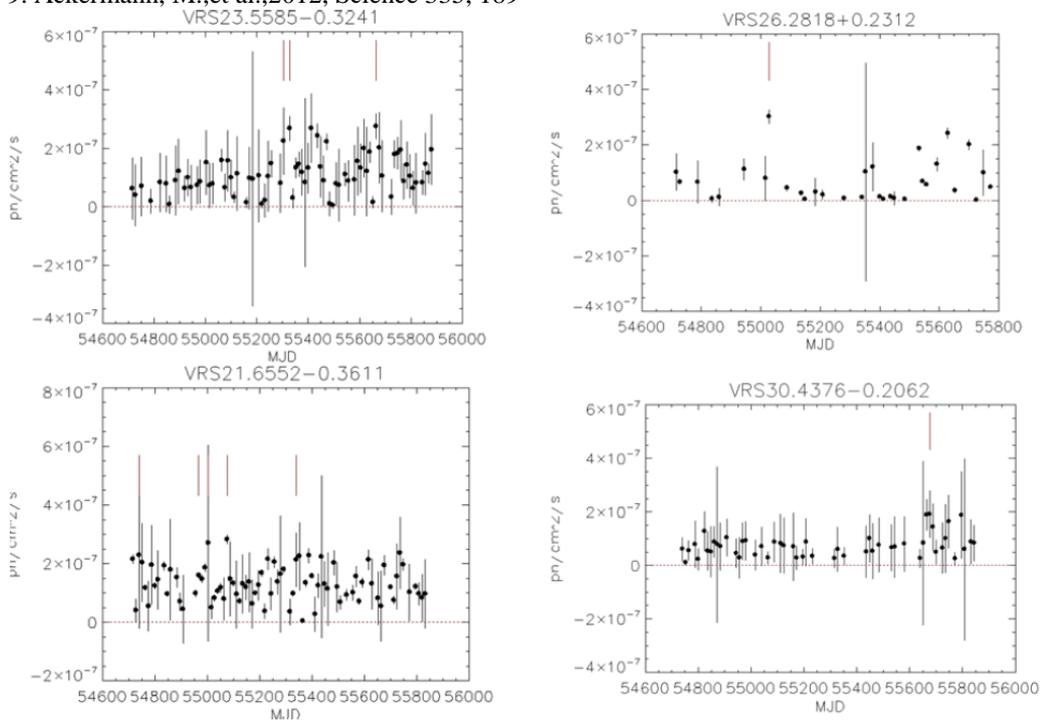

Figure 2. Hard X-ray light curves derived from archival INTEGRAL data. The red vertical lines indicate epochs where our Fermi analysis has indicated possible gamma-ray flaring. While not statistically significant there is a least a hint of associated hard-ray intensity increases in several cases.